\documentclass[aps,pra,amsmath,amssymb,twocolumn,showpacs]{revtex4-1} %tirei o twocolumn
\pdfoutput=1

\usepackage{graphicx}% Include figure files
\usepackage{dcolumn}% Align table columns on decimal point
\usepackage{bm}% bold math
\usepackage{bbold}% bold math

\newcommand{\ket}[1]{\ensuremath{|#1\rangle}}
\newcommand{\bra}[1]{\ensuremath{\langle #1|}}

\begin{document}

\preprint{PRE/003} 

\title{Entanglement properties of optical coherent states under amplitude damping}

\author{Ricardo Wickert}
 \email{ricardo.wickert@mpl.mpg.de}

\author{Nadja Kolb Bernardes}%
% \email{nadja.bernardes@mpl.mpg.de}

\author{Peter van Loock}%
% \email{peter.vanloock@mpl.mpg.de}

\affiliation{Optical Quantum Information Theory Group, Max Planck Institute for the Science of Light, G\"unther-Scharowsky-Str. 1/Bau 24, 91058 Erlangen, Germany}
\affiliation{Institute of Theoretical Physics I, Universit\"at Erlangen-N\"urnberg, Staudttr. 7/B2, 91058 Erlangen, Germany}

\date{\today}% It is always \today, today,

%\keywords{Concurrence, Entanglement, Coherent state superposition}

\begin{abstract}
Through concurrence, we characterize the entanglement properties of optical coherent-state qubits subject to an amplitude damping channel. We investigate the distillation capabilities of known error correcting codes and obtain upper bounds on the entanglement depending on the non-orthogonality of the coherent states and the channel damping parameter. This work provides a first, full quantitative analysis of these photon-loss codes which are naturally reminiscent of the standard qubit codes against Pauli errors.
\end{abstract}

\maketitle

\section{Introduction} %\label{sec:level1}

Entanglement has been established as a necessary resource for the implementation of many useful quantum primitives - teleportation of unknown states \cite{BennettTeleport}, key distribution \cite{BB84} and computational speed-up in classically exponential-time calculations \cite{Deutsch, Shor}, to cite but a few examples. This valued resource must, however, be protected against undesired interactions with the environment which lead, ultimately, to the decoherence of the quantum system. 

There are two proven ways of safeguarding the fragile quantum states from decohering: Quantum Error Correction (QEC) codes protect the information through encoding in a larger Hilbert space; Entanglement Purification (EP) protocols aim to distill entanglement from a number of identically prepared copies. It is known that QEC codes can be recast as EP schemes and vice-versa; the connection for discrete-variable (DV) systems was established in the seminal paper by Bennett \textit{et al} \cite{Bennett}, whereas the corresponding bridge for continuous-variables (CV) has been demonstrated in Refs. \cite{choi,Giedke,CerfNoGo}. 

In the optical context, amplitude damping - the absorption of transmitted photons - is a dominant source of decoherence. It can be appropriately modeled by having the signal interact with a vacuum mode in a beam splitter with the appropriately chosen transmissivity parameter. Amplitude damping appears thus as a \textit{Gaussian} error, and as such it is known that Gaussian resources - the set of operations which can easily be implemented through Gaussian ancilla, beam-splitters, phase-shifters and homodyne measurements - are of no use in protecting the signal state \cite{CerfNoGo,Giedke,Eisert,Fiurasek}, and more elaborate (and possibly non-deterministic) non-Gaussian operations must be accounted for.

In the present paper we review such a QEC scheme as proposed by Glancy \textit{et al.}\cite{GVR}, which protects arbitrary coherent-state superpositions (CSS, also known as ``cat states" \cite{RalphLund}) through the use of non-Gaussian encoding operations. We provide a hitherto inexistent quantitative analysis of this code's performance, examining its entanglement distillation capabilities through Wootters' concurrence \cite{Wooters}. This study is carried both directly and through the use of entanglement evolution equations \cite{konrad}, thus demonstrating their prowess over non-Gaussian CV carriers - even though the logical information may be codified in discrete qubits. 

Entangled coherent states subjected to an amplitude damping channel were also investigated by P. Munhoz \textit{et al.} with similar methods \cite{Munhoz}, however, employing a different class of states and aiming towards distinct applications from those presented in this contribution \footnote{In fact, the cluster-type entangled states of \cite{Munhoz} employ a bit-flip redundancy, whereas the amplitude damping channel manifests itself in the coherent-state basis as a phase-flip. The code employed here is designed with this particular error model in mind.}.

A coherent-state $ |\alpha \rangle$, defined as an eigenvalue of the annihilation operator ($\hat{a} |\alpha\rangle = \alpha |\alpha\rangle$), can be expressed in the Fock (number) basis as
\begin{equation}
|\alpha\rangle = e^{-|\alpha|^2} \sum_{n=0}^{\infty} \frac{\alpha^n}{\sqrt{n!}}|n\rangle \quad .
\end{equation}

A detailed introduction to the properties of coherent states can be found for instance in \cite{Mandel,Scully}.

\begin{figure}[th!]
\includegraphics[scale=0.50]{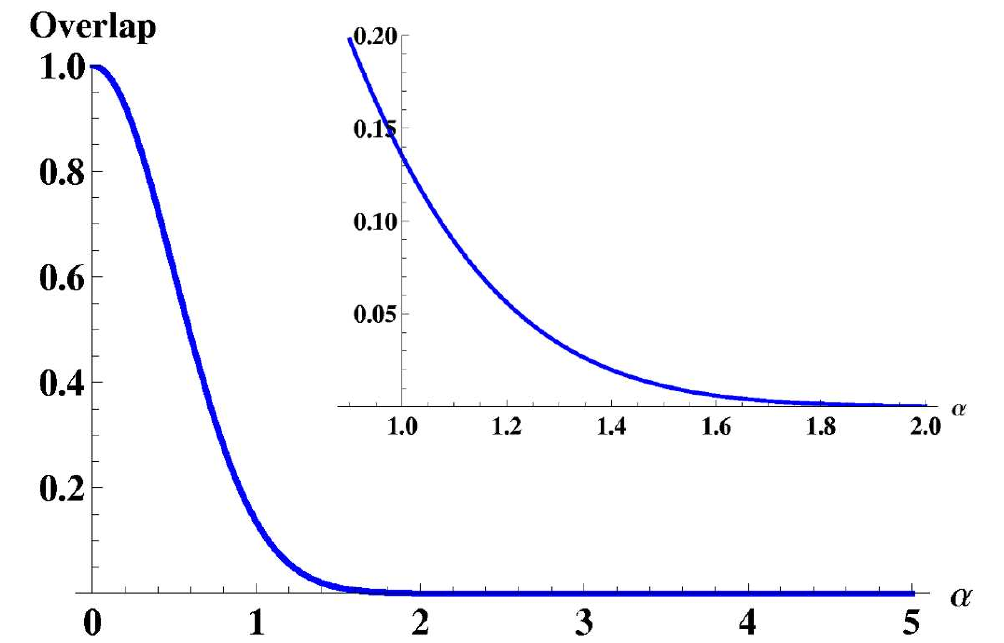}% Here is how to import EPS art
\caption{\label{fig:overlap} (Color online) Overlap between two coherent states, $\langle\alpha|-\alpha\rangle$ as a function of the size of the coherent state $|\alpha|$. Inset shows range accessible with current technology.}
\end{figure}

Following \cite{GVR}, we identify the logical qubits as $|0\rangle_L = |-\alpha\rangle$ and $|1\rangle_L = |\alpha\rangle$, in the so-called $(-,+)$ encoding. Qubits can equally be defined in the $(0,\alpha)$ encoding as $|0\rangle_L = |0\rangle$ and $|1\rangle_L = |2\alpha\rangle$, but it can be shown that both are equivalent in their decoherence properties and can easily be translated via displacement operations, so the first convention will be adopted here. An arbitrary qubit superposition is therefore represented as
\begin{equation}
|Q_\alpha\rangle = \frac{1}{\sqrt{N(\alpha)}} ( a |-\alpha\rangle + b |\alpha\rangle ) \quad 
\end{equation}
where $|a|^2 + |b|^2 = 1$ and $N(\alpha)$ is a normalization constant, $N(\alpha) = 1 + e^{-2|\alpha|^2}(ab^*+a^*b)$. It is argued that, for sufficiently large values of $\alpha$, $|-\alpha\rangle$ and $|\alpha\rangle$ are approximately orthogonal, and $N(\alpha) \approx 1$; however, present-day technologies only achieve limited $\alpha$ sizes (``Schr\"odinger Kittens" \cite{GrangierScience}). Therefore, a significant amount of non-orthogonality must be considered (fig. \ref{fig:overlap}).

\section{Amplitude Damping} %\label{sec:level1} 

Photon loss is considered to be the predominant source of errors to affect qubits in the optical context \cite{Yamamoto}. We model such loss by interacting the signal with a vacuum mode $|0\rangle_l$ in a beam splitter of transmissivity $\eta$, resulting in
\begin{eqnarray}
\label{bssup}
|Q\rangle_T &=& \frac{1}{\sqrt{N(\alpha)}} ( a |-\alpha\sqrt{\eta}\rangle|-\alpha\sqrt{1-\eta}\rangle_l + \nonumber \\
&& b |\alpha\sqrt{\eta}\rangle|\alpha\sqrt{1-\eta}\rangle_l )
\end{eqnarray}

The final state after transmission is obtained by integrating over the loss mode (denoted here by $|\beta\rangle_l$):
\begin{equation}
\label{trace}
\rho = \frac{1}{\pi}\int {} d^2\beta \,\: _l\langle\beta|Q\rangle_T {} _T\langle Q| \beta \rangle_l 
\end{equation}

\begin{figure}[h!]
\includegraphics[scale=0.50]{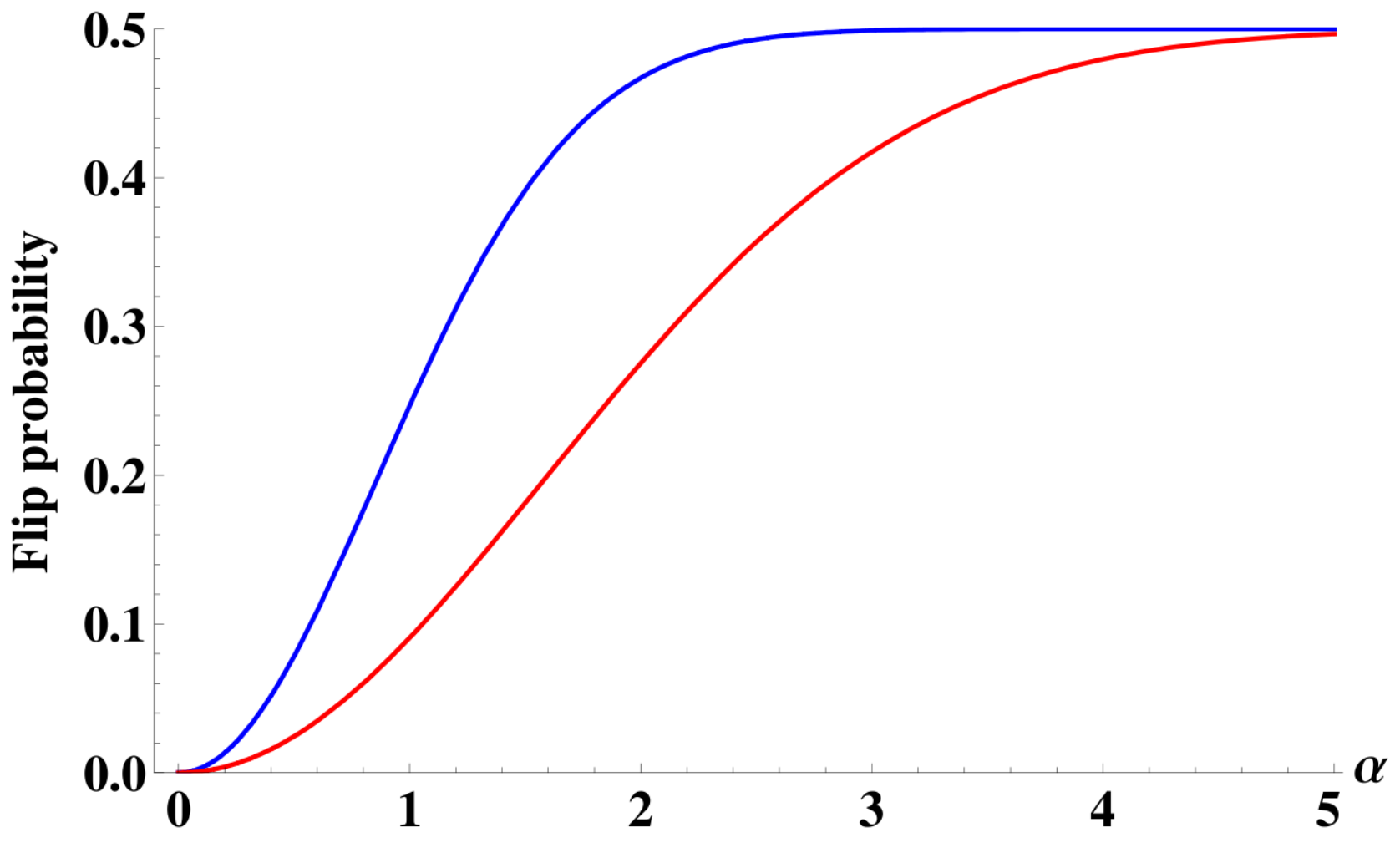}% Here is how to import EPS art
\caption{\label{fig:flipprob} (Color online) Phase flip probability $p_e$ as a function of the coherent state size $|\alpha|$, for channel transmissivities $\eta  = 2/3$ (blue line, above) and $\eta = 0.9$ (red line, below).}
\end{figure}

For a single coherent state, this integration is trivial and amounts to an amplitude contraction, remaining in a pure state. For a superposition, though, the resulting state after tracing out the loss mode is now mixed. One obtains
\begin{eqnarray}
\label{tracemixed}
\rho &=& (1-p_e)\frac{1}{\sqrt{N(\alpha)}}(a |-\alpha \sqrt{\eta}\rangle + b |\alpha \sqrt{\eta}\rangle ) \times H.c. \nonumber \\
&& + p_e \frac{1}{\sqrt{N^\prime(\alpha)}}(a |-\alpha \sqrt{\eta}\rangle - b |\alpha \sqrt{\eta}\rangle ) \times H.c. ,
\end{eqnarray}
where $H.c.$ is the Hermitian conjugate of the previous term and $N^\prime(\alpha) = 1 - e^{-2|\alpha|^2}(ab^*+a^*b)$. In order to simplify the analysis of the decohered state, Eq. (\ref{tracemixed}) can be cast into a more convenient form \cite{GVR}, namely,
\begin{equation}
\label{tracemixed2}
\rho = (1-p_e)|Q_{\alpha\sqrt{\eta}}\rangle \langle Q_{\alpha\sqrt{\eta}}| + p_e Z|Q_{\alpha\sqrt{\eta}}\rangle \langle Q_{\alpha\sqrt{\eta}}|Z
\end{equation}
where $p_e = \frac{1}{2}(1-e^{-2(1-\eta)|\alpha|^2})$ is the probability that the Pauli Z operator ($Z(a|0\rangle_L + b|1\rangle_L) = a|0\rangle_L - b|1\rangle_L$) was applied (fig. \ref{fig:flipprob}). With this expression, photon loss can be seen as having a two-fold effect: first, the amplitude of the states is unconditionally reduced from $\alpha$ to $\alpha \sqrt{\eta}$; second, with probability $p_e$, the qubit suffers a phase flip. 

\section{Error Correction}
%, as shown in fig. (\ref{fig:ecc}),
Having identified the effect of amplitude damping as a phase flip, a traditional 3-mode error-correcting code \cite{NielsenChuang} can be used to protect the qubit. Such a code can be implemented \cite{GVR} in the optical setting by sending the input state through a sequence of three beam-splitters followed by Hadamard gates - a highly non-Gaussian operation which implements, up to a normalization constant, $|0\rangle_L \rightarrow |0\rangle_L + |1\rangle_L$, $|1\rangle_L \rightarrow |0\rangle_L - |1\rangle_L$. The (unnormalized) encoded state which results is
\begin{equation}
\label{encoded}
a( |-\alpha\rangle + |\alpha\rangle )^{\otimes 3} + b( |-\alpha\rangle - |\alpha\rangle )^{\otimes 3}
\end{equation}

After transmission through the loss channels, another Hadamard gate is applied to each of the modes, which are then recombined through an inverted sequence of beam-splitters. The two ancilla modes are measured to provide syndrome information, from which the appropriate correcting operation can be applied to return the signal to its ``unflipped" state. Finally, teleporting the state into an appropriately prepared Bell state $|-\alpha\sqrt{\eta}\rangle|-\alpha\rangle + |\alpha\sqrt{\eta}\rangle|\alpha\rangle$, the amplitude can be restored to its original value.

The three-way redundant encoding achieved by the procedure outlined above can correct up to one error; therefore, the probability of achieving an error-free transmission is given by
\begin{equation}
\label{Psuccess}
p_{success,3} = 1 - 3 p_e^2 + 2 p_e^3 \quad .
\end{equation}
This can be increased by encoding the input state with a higher number of repetitions. We will thus also analyze codes with 5, 11 and 51 repetitions. 5-way redundancy increases the success probability to $1-10 p_e^3 + 15 p_e^4 - 6 p_e^5$; a n-repetition achieves
\begin{equation}
\label{pgeneral}
p_{success,n} = \sum_{k=0}^{\frac{n-1}{2}} \binom{n}{n-k} (1-p_e)^{n-k} p_e^k  \quad .
\end{equation}

\section{Entanglement}

The main focus on the QEC literature cited above lies on the achievement of the encodings - for instance, the implementation of the Hadamard gates or the teleportation strategy. The scheme's overall performance, though, is not quantified except for certain success probabilities or the fidelities of the involved operations. However, it has been noted that ``fidelity is insufficient to quantify quantum processes and protocols" \cite{bjork}. As such, we will explore the known fact that QEC codes can be recast as EP protocols \cite{Bennett,Aschauer} and, employing the entanglement as a figure of merit, provide quantitative benchmarks for this codification.
\subsection{Direct calculation}

We will consider initial cat states of the form
\begin{eqnarray}
&& |\chi_{\alpha_1,\alpha_2}\rangle = \\
&& \frac{1}{\sqrt{\tilde{N}(\alpha_1,\alpha_2)}} \left( \sqrt{w}|\alpha_1,\alpha_2\rangle + e^{i\theta}\sqrt{1-w}|-\alpha_1,-\alpha_2\rangle \right) \nonumber  
\end{eqnarray}
with $\tilde{N}(\alpha_1,\alpha_2) = 1 + 2 \cos{\theta} \sqrt{w(1-w)} e^{-2|\alpha_1|^2-2|\alpha_2|^2}$ and $|\alpha_1,\alpha_2\rangle$ a shorthand notation for $|\alpha_1\rangle|\alpha_2\rangle$.

In the case of direct transmission, the first mode is kept, while the second is sent through the photon-loss channel, resulting in
\begin{eqnarray}
\label{rhodirect}
\rho_{direct} &=& (1-P_e) |\chi_{\alpha,\alpha\sqrt{\eta}} \rangle \langle \chi_{\alpha,\alpha\sqrt{\eta}}| \nonumber \\
&& + P_e Z |\chi_{\alpha,\alpha\sqrt{\eta}} \rangle \langle \chi_{\alpha,\alpha\sqrt{\eta}}| Z \quad ,
\end{eqnarray}
with $P_e$ being the phase flip probability for a two-mode state, adjusted to preserve the normalization of each component, defined as
\begin{equation} P_e=\frac{1-e^{4\left|\alpha\right|^2}-e^{2\left|\alpha\right|^2(1-\eta)}+e^{2\left|\alpha\right|^2(1+\eta)}}{2(1-e^{4\left|\alpha\right|^2})}.
\end{equation}

If the sender and receiver make use of the 3-mode repetition, the entangled pair they will share after encoding, transmission and decoding is given by
\begin{eqnarray}
\label{rhoenc3}
\rho_{final,3} && = (1 - 3 P_e^2 + 2 P_e^3) |\chi_{\alpha,\alpha\sqrt{\eta}} \rangle \langle \chi_{\alpha,\alpha\sqrt{\eta}}| \nonumber \\
&& + (3 P_e^2 - 2 P_e^3) Z |\chi_{\alpha,\alpha\sqrt{\eta}} \rangle \langle\chi_{\alpha,\alpha\sqrt{\eta}}| Z  ,
\end{eqnarray}
with the $Z$ operator taken to act, in both (\ref{rhodirect}) and (\ref{rhoenc3}), on the transmitted mode. The use of higher order n-repetitions results in
\begin{eqnarray}
\label{rhoencn}
\rho_{final,n} && = P_{success,n} |\chi_{\alpha,\alpha\sqrt{\eta}} \rangle \langle \chi_{\alpha,\alpha\sqrt{\eta}}| \\
&& + (1 - P_{success,n}) Z |\chi_{\alpha,\alpha\sqrt{\eta}}\rangle \langle\chi_{\alpha,\alpha\sqrt{\eta}}| Z .\nonumber
\end{eqnarray}
with $P_{success,n}$ defined as in (\ref{pgeneral}), but depending on $P_e$ instead of $p_e$.

We adopt Wootters' concurrence to quantify the entanglement; in the case of a bipartite qubit system, the concurrence is given by \cite{Wooters}
\begin{equation}
\label{conceq}
C = max\{0,\sqrt{\lambda_1}-\sqrt{\lambda_2}-\sqrt{\lambda_3}-\sqrt{\lambda_4} \}
\end{equation}
where $\lambda_i$ are the eigenvalues, listed in decreasing order, of $\rho \tilde{\rho}$. $\tilde{\rho}$ is the time-reversed density operator,
\begin{equation}
\label{magictr}
\tilde{\rho} = ( \sigma_{y,1} \otimes \sigma_{y,2} ) \rho^{*} ( \sigma_{y,1} \otimes \sigma_{y,2} ) \quad ,
\end{equation}
where $\sigma_{y,i}$ is the Pauli $Y$ operator in the $i$-th mode. In order to cast the density matrix in the appropriate form, we make use of an orthogonal basis $\{ |u_{\alpha}\rangle,|v_{\alpha}\rangle \}$ (see for instance \cite{wang}) such that
\begin{align}
|\alpha\rangle = \mu_\alpha |u_{\alpha}\rangle& + \nu_\alpha |v_{\alpha}\rangle \\
|-\alpha\rangle = \mu_\alpha |u_{\alpha}\rangle& - \nu_\alpha |v_{\alpha}\rangle \nonumber \\
\mbox{with } \mu_\alpha = \left(\frac{1+e^{-2|\alpha|^2}}{2}\right)^{\frac{1}{2}}& \quad \textrm{and} \quad \nu_\alpha = \left(\frac{1-e^{-2|\alpha|^2}}{2}\right)^{\frac{1}{2}} \nonumber
\end{align}
This way, the entangled pair can be written, up to a normalization factor of $1/\sqrt{\tilde{N}(\alpha,\alpha)}$ (omitted below for clarity), as

\begin{align}
|\chi_{\alpha,\alpha}\rangle &= \sqrt{w}|\alpha\rangle|\alpha\rangle + e^{i\theta}\sqrt{1-w}|-\alpha\rangle|-\alpha\rangle  \\
&= (\sqrt{w} + e^{i\theta}\sqrt{1-w}) \left( \mu_\alpha^2 |u_{\alpha}\rangle|u_{\alpha}\rangle + \nu_\alpha^2 |v_{\alpha}\rangle|v_{\alpha}\rangle \right) \nonumber \\
&\ +(\sqrt{w} - e^{i\theta}\sqrt{1-w}) \mu_\alpha \nu_\alpha \left( |u_{\alpha}\rangle|v_{\alpha}\rangle + |v_{\alpha}\rangle|u_{\alpha}\rangle \right) \nonumber 
\end{align}
and thus
\begin{equation}
\label{matrixunflip}
|\chi_{\alpha,\alpha}\rangle \langle \chi_{\alpha,\alpha}| = \left(
\begin{array}{cccc}
|a|^2 & ab^* & ac^* & ad^* \\
ba^* & |b|^2 & bc^* & bd^* \\
ca^* & cb^* & |c|^2 & cd^* \\
da^* & db^* & dc^* & |d|^2
\end{array}
\right)
\end{equation}
where 
\begin{eqnarray}
a &=& (\sqrt{w}+e^{i\theta}\sqrt{1-w})\mu_\alpha^2 \nonumber \\
b = c &=& (\sqrt{w}-e^{i\theta}\sqrt{1-w})\mu_\alpha \nu_\alpha \nonumber \\
d &=& (\sqrt{w}+e^{i\theta}\sqrt{1-w})\nu_\alpha^2  \quad .\nonumber
\end{eqnarray}

After transmission, a similar matrix is also obtained for the unflipped state ($|\chi_{\alpha,\alpha\sqrt{\eta}}\rangle$); the phase-flipped states after transmission are described by
\begin{equation}
\label{matrixflip}
Z |\chi_{\alpha,\alpha\sqrt{\eta}}\rangle \langle \chi_{\alpha,\alpha\sqrt{\eta}} | Z = \left(
\begin{array}{cccc}
|\tilde{a}|^2 & \tilde{a}\tilde{b}^* & \tilde{a}\tilde{c}^* & \tilde{a}\tilde{d}^* \\
\tilde{b}\tilde{a}^* & |\tilde{b}|^2 & \tilde{b}\tilde{c}^* & \tilde{b}\tilde{d}^* \\
\tilde{c}\tilde{a}^* & \tilde{c}\tilde{b}^* & |\tilde{c}|^2 & \tilde{c}\tilde{d}^* \\
\tilde{d}\tilde{a}^* & \tilde{d}\tilde{b}^* & \tilde{d}\tilde{c}^* & |\tilde{d}|^2
\end{array}
\right)
\end{equation}
with $Z$ taken to act on mode 2, and 
\begin{eqnarray}
\tilde{a} &=& (\sqrt{w}-e^{i\theta}\sqrt{1-w})\mu_\alpha\mu_{\alpha\sqrt{\eta}}\nonumber \\
\tilde{b} &=& (\sqrt{w}+e^{i\theta}\sqrt{1-w})\mu_\alpha \nu_{\alpha\sqrt{\eta}}\nonumber \\
\tilde{c} &=& (\sqrt{w}+e^{i\theta}\sqrt{1-w})\mu_{\alpha\sqrt{\eta}}\nu_\alpha\nonumber \\
\tilde{d} &=& (\sqrt{w}-e^{i\theta}\sqrt{1-w})\nu_\alpha\nu_{\alpha\sqrt{\eta}}\quad .\nonumber
\end{eqnarray}

\begin{figure}[h!]
\includegraphics[scale=0.50]{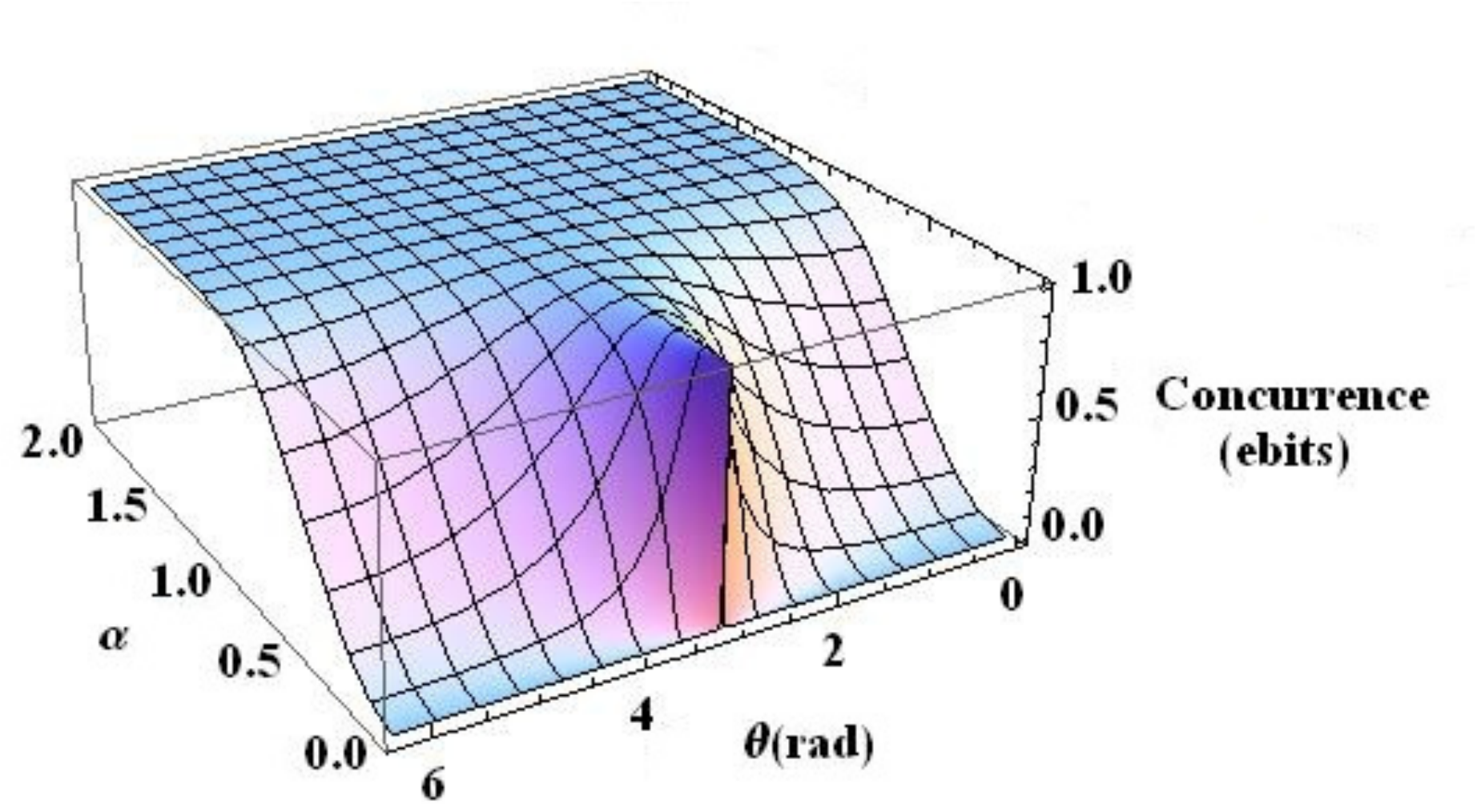}% Here is how to import EPS art
\caption{\label{fig:blabla3d} (Color online) Concurrence for the initial state as a function of the size of the coherent state $|\alpha|$ and the phase parameter $\theta$.}
\end{figure}

With (\ref{matrixunflip}) and (\ref{matrixflip}), one can construct the matrices (\ref{rhodirect}), (\ref{rhoenc3}) or (\ref{rhoencn}) and, through the use of (\ref{magictr}), calculate the concurrence as defined in (\ref{conceq}). Fig. \ref{fig:blabla3d} plots the entanglement of the initial state with $w=1/2$, corresponding to ``genuine" Bell states. It is seen that $\theta=0$ and $\theta=\pi$ (respectively ``even" and ``odd" due to the parity of the number states in the corresponding superposition) result in maximum entanglement, the former asymptotically for large superpositions, the latter independently of the superposition size. %For any value of $\alpha$, the concurrence is zero when $\theta=\pi/2$ or $\theta=3\pi/2$, as noted in \cite{vanenk}. 

The above method, however, is not convenient for calculating the entanglement of different levels of encoding, as it requires the eigenvalues to be computed for each variation, which can be a time-consuming task if done analytically. Thus we present, in the following section, an alternative way to obtain the state's concurrence.

\subsection{Entanglement Evolution}

The effect of an arbitrary quantum channel $\$$ to a state can also be described in the dual picture \cite{choi}, interchanging the roles of the channel and the initial state. An evolution equation is obtained \cite{konrad}, which equates the entanglement (and in particular the concurrence) of the final state to the product of the concurrence of a maximally entangled state $\ket{\phi^{+}}$ subjected to the same channel times the concurrence of the initial state $\ket{\chi}$. The problem is thus reduced to the calculation of two, possibly simpler, concurrences.
\begin{equation}
	C\left[\left(1\otimes\$\right)\ket{\chi}\bra{\chi}\right] = C\left[\left(1\otimes\$\right)\ket{\phi^{+}}\bra{\phi^{+}}\right]C\left[\ket{\chi}\right].
\label{ent.evo}
\end{equation}

As in the previous section, the states will be written in the orthogonal basis $\{\ket{u_{\alpha}},\ket{v_{\alpha}}\}$. One possible maximally entangled Bell state in this basis is given as $\ket{\phi^{+}_{\alpha,\alpha}}=\left(\ket{u_\alpha}\ket{u_\alpha}+\ket{v_\alpha}\ket{v_\alpha}\right)/\sqrt{2}$, but any state carrying exactly one ebit will have its entanglement affected in the same way. The resulting state after transmission through the amplitude damping channel can be described by an X matrix 
\begin{equation}
\left(1\otimes\$\right)\ket{\phi^{+}_{\alpha,\alpha}}\bra{\phi^{+}_{\alpha,\alpha}}=
\left(
\begin{array}{cccc}
a & 0 & 0 & f \\
0 & b & z & 0 \\
0 & z* & c & 0 \\
f* & 0 & 0 & d \\
\end{array} \right),
\label{matx}
\end{equation}
where
\begin{eqnarray}
a &=&  \frac{(1+e^{-2\left|\alpha\right|^2(1-\eta)})\mu^2_{\tiny{\alpha\sqrt{\eta}}}}{4\mu_{\alpha}^2} \nonumber \\
b &=& -\frac{(-1+e^{-2\left|\alpha\right|^2(1-\eta)})\nu^2_{\tiny{\alpha\sqrt{\eta}}}}{4\mu_{\alpha}^2}  \nonumber \\
c &=&  \frac{-(-1+e^{-2\left|\alpha\right|^2(1-\eta)})\mu^2_{\tiny{\alpha\sqrt{\eta}}}}{4\mu_{\alpha}^2}  \nonumber \\
d &=&  \frac{(-1+e^{-2\left|\alpha\right|^2(1-\eta)})\mu_{\tiny{\alpha\sqrt{\eta}}}\nu_{\tiny{\alpha\sqrt{\eta}}}}{4\mu_{\alpha}\nu_{\alpha}} \nonumber \\
f &=& -\frac{(1+e^{-2\left|\alpha\right|^2(1-\eta)})\mu_{\tiny{\alpha\sqrt{\eta}}}\nu_{\tiny{\alpha\sqrt{\eta}}}}{4\mu_{\alpha}\nu_{\alpha}} \nonumber \\
z &=& \frac{(-1+e^{-2\left|\alpha\right|^2(1-\eta)})\mu_{\tiny{\alpha\sqrt{\eta}}}\nu_{\tiny{\alpha\sqrt{\eta}}}} {4\mu_{\alpha}\nu_{\alpha}} \nonumber .
\end{eqnarray}

The outer and inner elements of the (\ref{matx}) represent, respectively, ``unflipped" and ``flipped" Bell states of reduced, $\alpha\sqrt{\eta}$, amplitude. Just as in (\ref{tracemixed}), this allows one to rewrite the resulting state as
\begin{eqnarray}
\left(1\otimes\$\right)\ket{\phi^{+}_{\alpha,\alpha}}\bra{\phi^{+}_{\alpha,\alpha}}&=&
(1-P_e) \ket{\phi^{+}_{\alpha,\alpha\sqrt{\eta}}}\bra{\phi^{+}_{\alpha,\alpha\sqrt{\eta}}} \nonumber \\
&& + P_e Z\ket{\phi^{+}_{\alpha,\alpha\sqrt{\eta}}}\bra{\phi^{+}_{\alpha,\alpha\sqrt{\eta}}}Z
\label{bellflipanaoflipa} \quad ,
\end{eqnarray}

Equation (\ref{ent.evo}) can also be extended to encoded states, such as those obtained by the QEC repetition codes described in Sec. III. In this case, the channel $\$$ will include not only the lossy transmission channels itself, but also the encoding, syndrome measurement, error correction and decoding operations. Nevertheless, the resulting density matrix for the Bell state is still an X matrix, 
\begin{eqnarray}
\left(1\otimes\$_{enc}\right)\ket{\phi^{+}_{\alpha,\alpha}}\bra{\phi^{+}_{\alpha,\alpha}} = P_{success,n}\ket{\phi^{+}_{\alpha,\alpha\sqrt{\eta}}}\bra{\phi^{+}_{\alpha,\alpha\sqrt{\eta}}} \nonumber \\
 +(1-P_{success,n})Z\ket{\phi^{+}_{\alpha,\alpha\sqrt{\eta}}}\bra{\phi^{+}_{\alpha,\alpha\sqrt{\eta}}}Z \quad,
\label{mat.bell.final}
\end{eqnarray}
where $P_{success,n}$, as in the Sec. III, is the probability of achieving error-free transmission.

The concurrence of a state described by an X matrix can easily be found by \cite{eberly}
\begin{equation}
	C\left(\rho\right) = 2 \mbox{ max} \left[0,|z|-\sqrt{ad},|f|-\sqrt{bc}\right].
\label{conX}
\end{equation}

The only remaining step is the calculation of the second term of the RHS of (\ref{ent.evo}). Through the methods outlined in the previous subsection, one finds: %(fig. \ref{fig:blabla3d})

\begin{equation}
 C\left[\ket{\chi}\right]=
\frac{2 \left( 1 - e^{-4 \left| \alpha \right|^2} \right) \sqrt{w(w-1)} }{1 + 2 \sqrt{w(w-1)} e^{-4 \left| \alpha \right|^2} \cos \theta} \; ,
\label{con.state}
\end{equation}

This method matches the results found in the previous subsection, but allows for a computationally significant speed-up in calculation times. The concurrences for different encodings are compared below:

\begin{figure}[h!]
\includegraphics[scale=0.4]{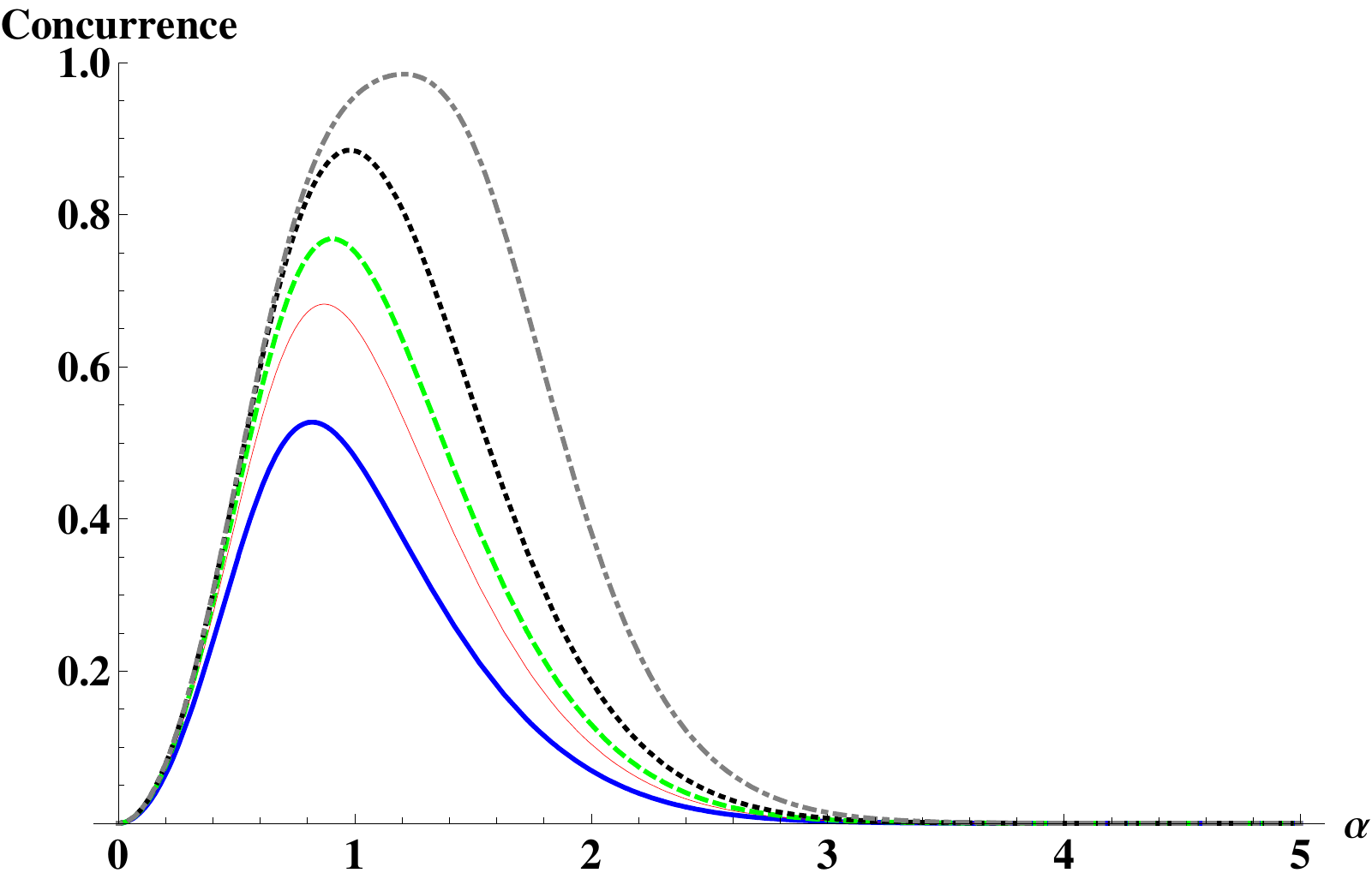} \\
(a) $\eta  = 2/3$, $\theta=0$ \\
\includegraphics[scale=0.4]{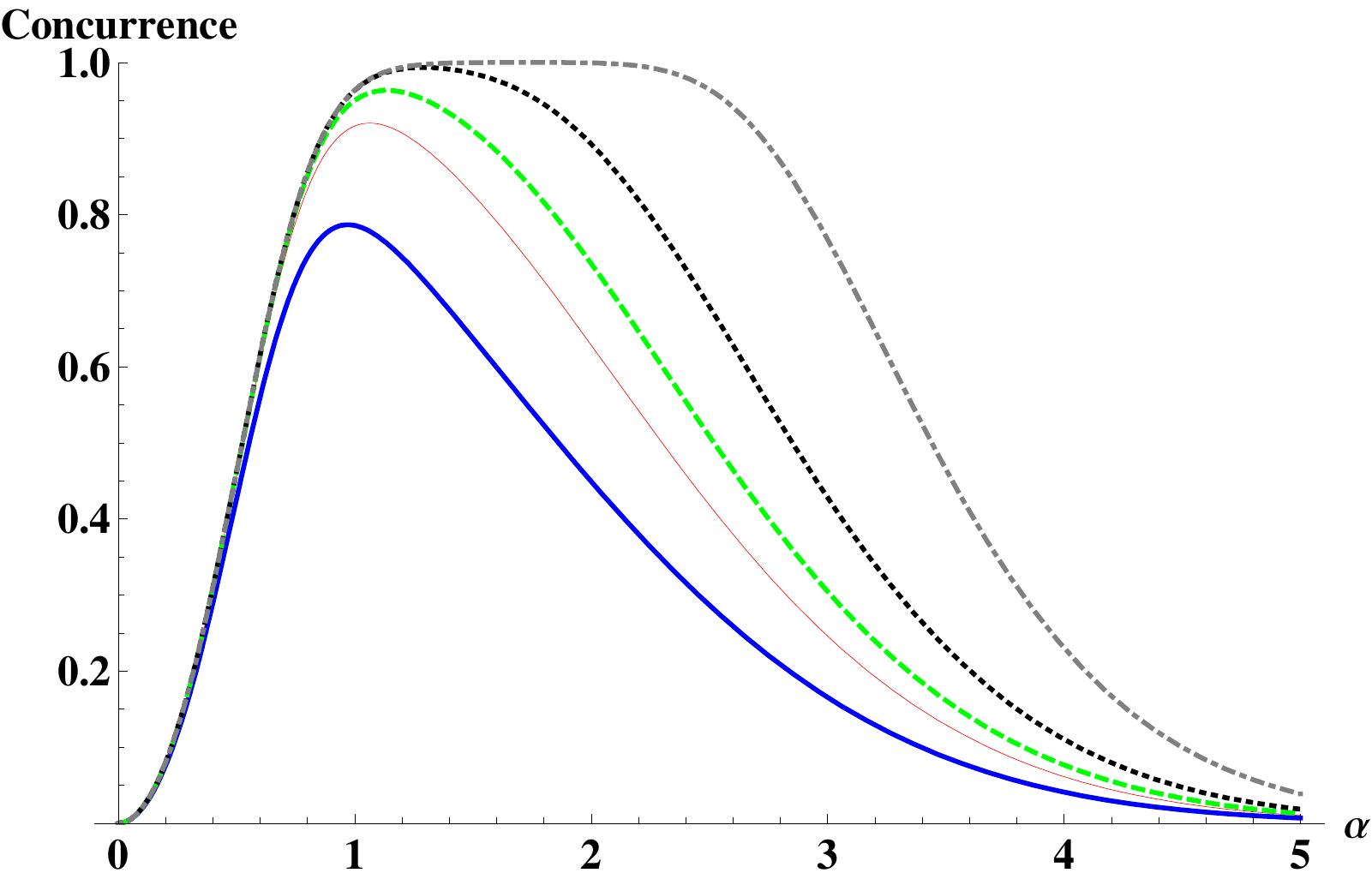} \\
(b) $\eta  = 0.9$, $\theta=0$ \\
\includegraphics[scale=0.4]{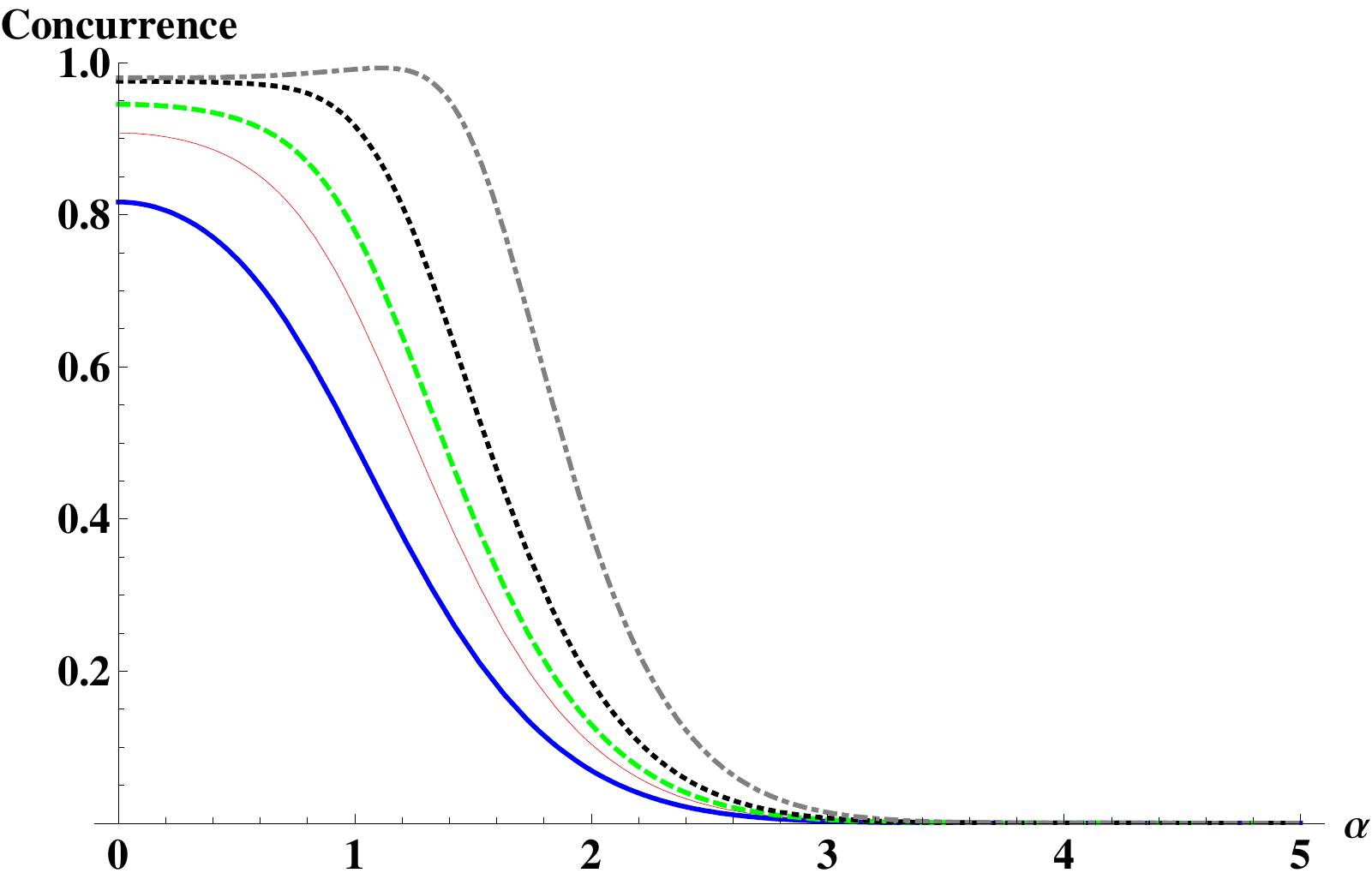} \\
(c) $\eta  = 2/3$, $\theta=\pi$ \\
\includegraphics[scale=0.4]{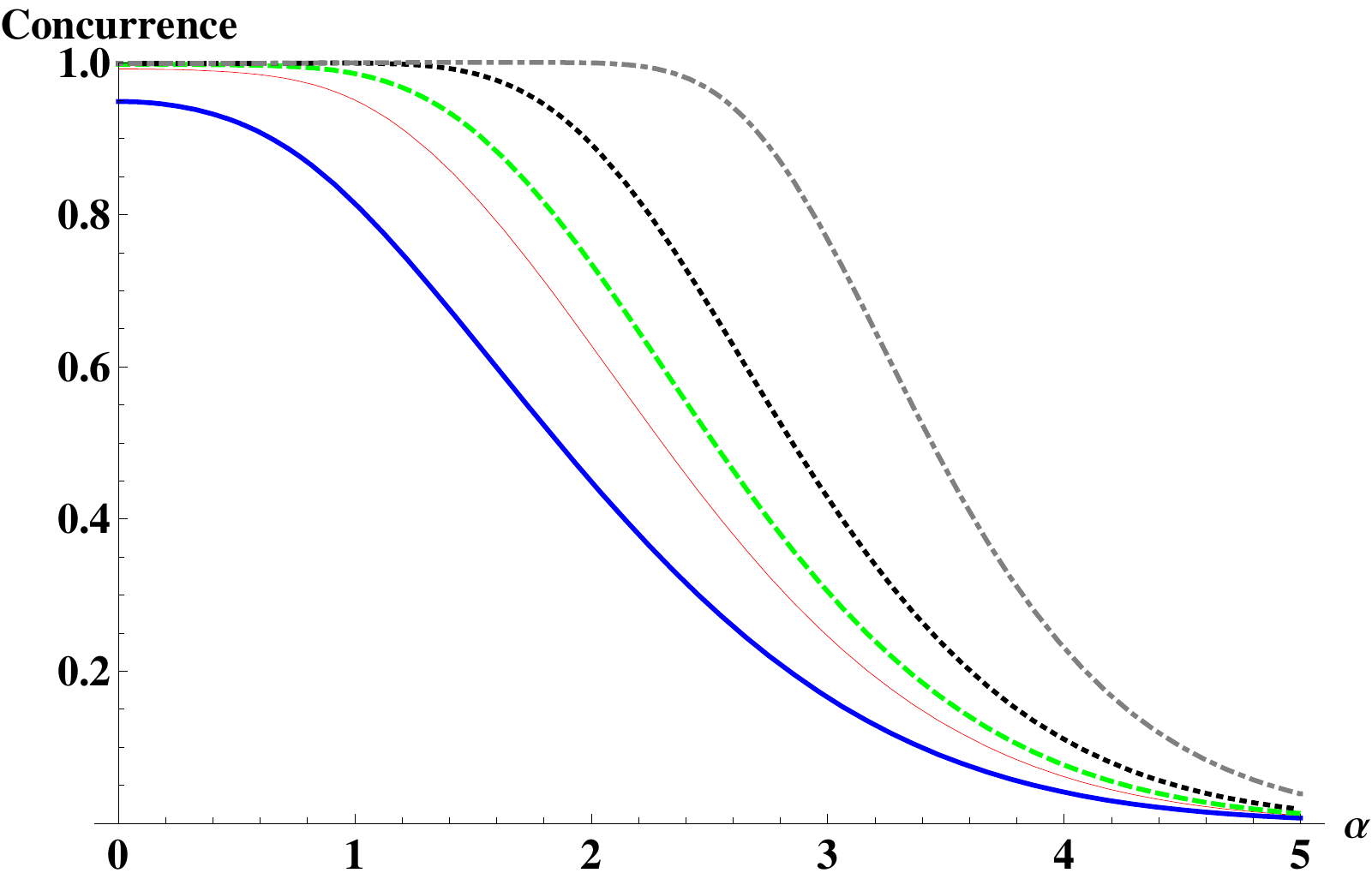} \\
(d) $\eta  = 0.9$, $\theta=\pi$% Here is how to import EPS art
\caption{\label{fig:condut1} (Color online) Concurrence after transmission  as a function of the coherent state size $|\alpha|$, for different values of transmissivity $\eta$ and the phase parameter $\theta$. Direct transmission (blue, thick line) is compared to encoding with 3 (red line), 5 (green, dashed), 11 (black, dotted) and 51 (grey, dash-dotted) qubits.}
\end{figure}

Two distinct observations can be made from fig. \ref{fig:condut1}. First, for the even cat states (a) and (b), the non-orthogonality of the basis states prevents the pair from achieving high entanglement for low values of $\alpha$; higher encodings are of little advantage in this regime. The choice of the phase $\theta$ is of paramount importance when one compares to the odd states (c) and (d), where even for very small $\alpha$ there is still a significant amount of shared entanglement, and encoding in more qubits still improves on this amount. Second, for sufficiently large sizes of $\alpha$, the flip probability approaches $0.5$ and thus it dephases the qubit entirely, independently of the phase or encoding. However, for a certain range, the different codifications are seen to help achieve and sustain higher entanglement between the shared pairs.

\begin{figure}[h!]
\includegraphics[scale=0.41]{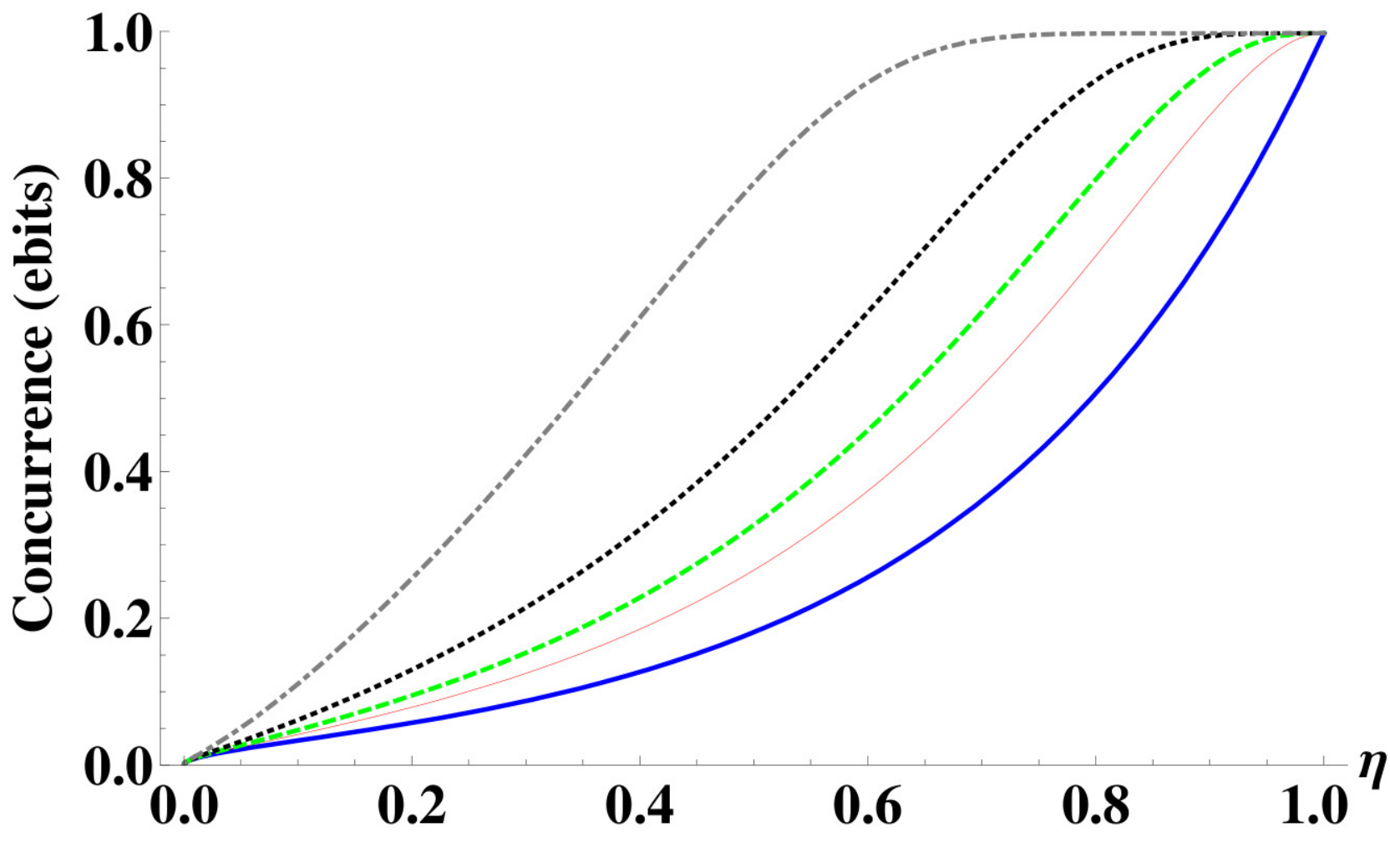}% Here is how to import EPS art
\caption{\label{fig:concfix15}(Color online) Concurrence after transmission as a function of the channel transmissivity $\eta$, shown without encoding (blue, thick line) and encoded with 3 (red line), 5 (green, dashed), 11 (black, dotted) and 51 (grey, dash-dotted) qubits. Here $\alpha  = 1.3$ and $\theta=0$. The behaviour for $\theta=\pi$ follows qualitatively the same form.}
\end{figure}

Finally, in fig. \ref{fig:concfix15} one can observe that, all the way to zero transmissivity, the pair has some residual entanglement. Higher redundancy improves this figure; we also note that, in the context of \cite{bjork}, the encoding does not induce entanglement sudden death (ESD) \cite{eberly}.

\section{Discussion and Conclusions} %\label{sec:level1} use of the code 
\begin{figure}[h!]
\includegraphics[scale=1.05]{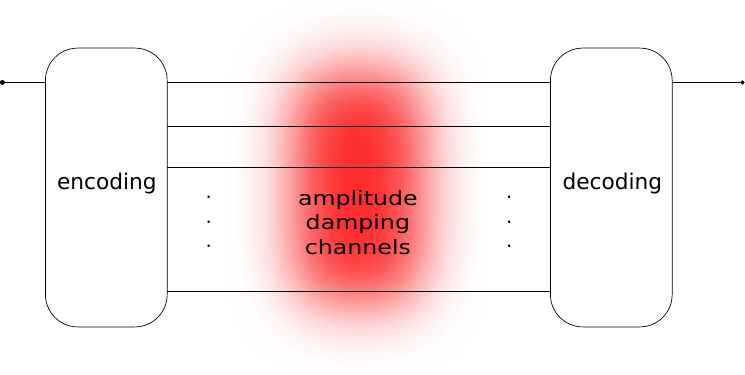} \\% Here is how to import EPS art \\
\includegraphics[scale=1.05]{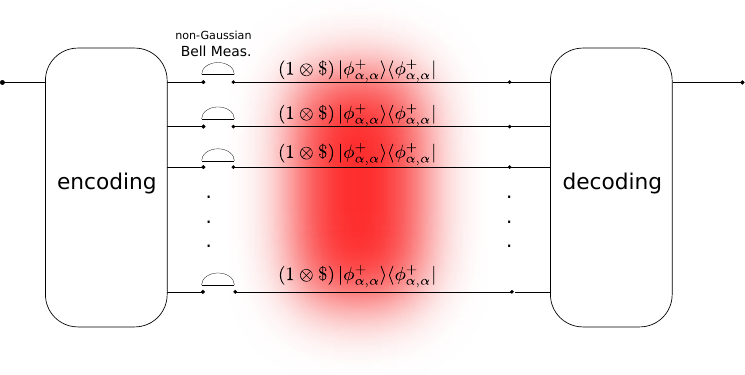}% Here is how to import EPS art
\caption{\label{fig:pdfeppecc} (Color online) The QEC code as presented in \cite{GVR} (above) and a possible interpretation as an EPP (below).}
\end{figure}

In this work, we have studied the decoherence process of entangled coherent-state superpositions in the amplitude damping channel, exploring QEC codes originated from the discrete-variable regime in an optical, continuous-variable setting. The quantitative analysis was reminiscent of EP protocols, however, this translation must be carried out carefully: a true distillation would, in addition to the non-Gaussian Hadamard gates necessary for the encoding, require teleportations to be performed via non-Gaussian two-mode Bell measurements \cite{Bennett,vanenk}. In this picture, multiple copies of the entangled resource $\ket{\phi^{+}}$ are shared across the amplitude damping channel. The use of the encoding and decoding circuits of the original error-correcting code then results in one pair with higher entanglement (see fig. \ref{fig:pdfeppecc}), which effectively characterizes this scheme as a one-way, deterministic entanglement distillation for noisy, non-Gaussian CV states.

Alternatively, in the context of \cite{Aschauer}, the translation to an entanglement purification protocol could be performed through an elaborate non-Gaussian multi-mode projective measurement in the encoding basis. The non-Gaussian elements are, in any case, a required condition, enabling the circumvention of known No-Go theorems \cite{CerfNoGo,Giedke,Eisert,Fiurasek}.

The protection granted against phase-flip (or $Z$) errors by different repetition codes was investigated and quantified through concurrence; however, possible physical limitations in the implementation of the gates necessary for these encodings have not been been taken into account, therefore establishing the above results as an upper bound. In particular, we note that both Hadamard and teleportation operations in \cite{GVR} only approach unit fidelity and/or high success probabilities for large superposition sizes. However, as we showed here, in this regime these codes would cease to work. We conclude that, in order to incorporate the actual encoding operations into the above protocol, gates that succeed with high fidelity on not too-large $\alpha$ cat states will be needed. One possibility to achieve this could be the scheme of \cite{RalphLund}, where $\alpha \approx 1.2$ is enough for performing fault-tolerant quantum computing. 

We have observed a trade-off between the orthogonality of the coherent-state basis and the probability of a phase-flip error. The former imposes a necessity for a maximum size of coherent-state superposition, while the latter prevents the use of arbitrarily large superpositions, thus hinting at an optimal regime depending on channel parameters. 

The generation of optical cat states remains an experimental challenge, though much progress has been recently achieved. Arbitrarily large squeezed cat states can be obtained through photon number states and homodyne detection \cite{Ourjoumtsev}, however such technique is highly probabilistic and presents only moderate output fidelities. A different approach involves tapping squeezed vacuum in a beam splitter (BS), with one output mode directed to a number-resolving photon counter. Conditional on the number of photons detected, the other mode is projected into an odd or even cat state \cite{GrangierScience, Takahashi}; an n-photon subtraction is described as $\hat{a}^n\hat{S}(\epsilon_0)|0\rangle$, where $\hat{a}$ is the annihilation operator, $\hat{S}$ is the squeezing operator and $\epsilon_0$ amounts to the degree of squeezing applied. Superpositions with $\alpha \sim 1.2-1.3$ can be created through this technique; it is also worth noting that the more ``valuable", 1-ebit odd CSS resource is more readily obtained than its less entangled even state counterpart.%. as the first requires the detection of one sole photon.%through photon subtraction of a squeezed coherent state ; The photon subtraction technique, described as

The choice of ``odd" cat states (with $\theta=\pi$) could be seen as a viable alternative to employ small-amplitude superpositions to distribute entanglement while minimizing the amplitude damping effects. Still, $\alpha$ may not be chosen too small so that the resulting qubits become impractical for information transfer or computational purposes. Nonetheless, in the context of optical, realistic quantum communication, a combination of ``odd" CSS resources, linear optics and nonlinear measurements for encoding may lead to efficient error correction and entanglement distillation strategies.

$\quad$

\subsection*{Acknowledgments}
Support from the Emmy Noether Program of the Deutsche Forschungsgemeinschaft is gratefully acknowledged.

\bibliography{labibio}

\end{document}